\documentclass[aps, prb, reprint, showpacs, floatfix]{revtex4-1} 
\usepackage{graphicx}
\usepackage{amsmath}
\usepackage{amssymb}

\begin{document}
\title{Quantum transport in graphene Hall bars: Effects of side gates}
\author{M. D. Petrovi\'c}\email{marko.petrovic@uantwerpen.be}
\affiliation{Department of Physics, University of Antwerp,\\
             Groenenborgerlaan 171, B-2020 Antwerp, Belgium}
\author{F. M. Peeters}\email{francois.peeters@uantwerpen.be}
\affiliation{Department of Physics, University of Antwerp,\\
             Groenenborgerlaan 171, B-2020 Antwerp, Belgium}
\begin{abstract}

Quantum electron transport in side-gated graphene Hall bars is investigated in
the presence of quantizing external magnetic fields. The asymmetric potential
of four side-gates distorts the otherwise flat bands of the relativistic
Landau levels, and creates new propagating states in the Landau spectrum
(i.e.~snake states). The existence of these new states leads to an interesting
modification of the bend and Hall resistances, with new quantizing plateaus
appearing in close proximity of the Landau levels. The electron guiding in
this system can be understood by studying the current density profiles of the
incoming and outgoing modes. From the fact that guided electrons fully
transmit without any backscattering (similarly to edge states), we are able to
analytically predict the values of quantized resistances, and they match
the resistance data we obtain with our numerical (tight-binding) method. These
insights in the electron guiding will be useful in predicting the resistances
for other side-gate configurations, and possibly in other system geometries,
as long as there is no backscattering of the guided states.

\end{abstract}
\maketitle{}
\section{Introduction}

Quantum Hall measurements~\cite{graphene_qhe_zhang, novoselov_qhe} in
graphene~\cite{novoselov_graphene} revealed the relativistic nature of its
charge carriers and the gapless spectrum. Long before the discovery of
graphene, it was known that carriers in a conventional two-dimensional
electron gas (2DEG) tend to move along snake like paths when exposed to
inhomogeneous magnetic fields, the so called {\em snake
states.\/}\cite{muller_snake_states, reijniers} Similar effects were explored
even earlier in the studies of electron propagation on the boundary of
magnetic domains in metallic systems.~\cite{rozhavski, cabrera1, cabrera2,
zakharov} Experiments in non-planar 2DEG\cite{bykov} and in systems with
ferromagnetic stripe\cite{nogaret_prl} indirectly measured the effects of
snake states on the longitudinal and the Hall resistances. In graphene,
theoretical predictions of snake-state effects~\cite{oroszlany} were quickly
followed by experiments which confirmed their existence.~\cite{williams_qhe}
Snake states in a Hall bar geometry were previously studied in
Ref.~\onlinecite{slavisa_hall} using a classical billiard model. A top-gate
was used to create a $pn$ junction along the main diagonal of the Hall cross,
and oscillations of the bend resistance were connected with electron guiding
along the snake-like paths at the $pn$ interface.

\begin{figure}[b]
\includegraphics{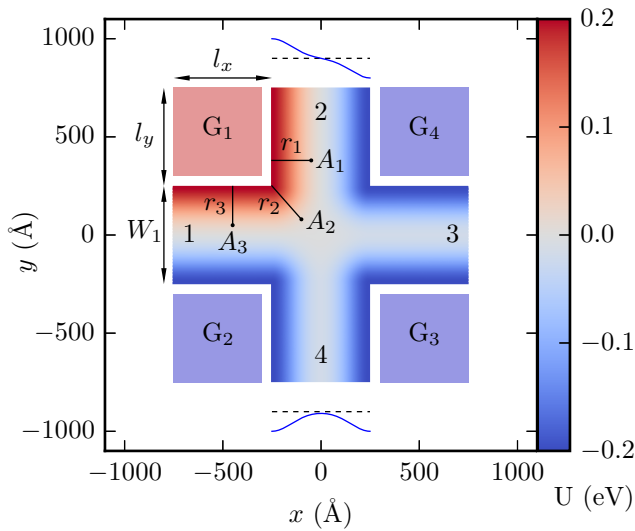}
\caption{\label{fig_system}
  (Color online) Side-gated Hall bar system. Four leads are marked with
  numbers, and they all have approximately equal width
  ($W_{1}=W_{3}=49.71$\,nm, and $W_{2}=W_{4}=49.94$\,nm, $l_x=l_y=50$\,nm).
  Horizontal leads have zigzag edges, while vertical leads have armchair
  edges, and all dangling bonds are removed. Three lines $r_1$, $r_2$, and
  $r_3$ show the minimal distances from the system edge to three arbitrary
  points $A_1$, $A_2$, $A_3$, respectively. These distances are used in
  Eq.~(\ref{eq_pot}) to calculate the gate potential of the $1^{\rm st}$ lead at the
  corresponding points $A_1$, $A_2$, and $A_3$. The potential profiles inside
  the leads are depicted by the blue curves above the $2^{\rm nd}$ and below 
  the $4^{\rm th}$ lead (black dashed lines mark the position of the zero
  potential).} \end{figure}

In this paper we study quantum transport of electrons in graphene Hall bars
surrounded with four side-gates (see Fig.~\ref{fig_system}). The gates modify
the local electron density on the edges of the Hall bar, and induce a local
electric potential. If the system is placed in an external magnetic field,
this edge potential guides the charge carriers along specific equipotential
lines. For weak fields, these states move along the previously mentioned
snake-like paths, while for stronger fields, we prefer to call them guided
states. Our main goal is to understand how this guiding occurs locally, and
which paths the electron takes inside the system. Our second goal is to
predict the experimentally measurable effects of this guiding. We start by
investigating how the side-gate potential modifies the dispersion relations of
the electrons in each lead. By studying the current density profiles of the
incoming and outgoing states in two representative leads, we are able to build
a physical picture of electron transport in this inhomogeneous system. This
picture, in combination with the Landauer-B{\"u}ttiker formalism, allows us to
analytically predict the quantization of the bend and Hall resistances. The
quantized resistance values match the ones that we obtain with our numerical,
tight-binding method. Although we choose one specific potential configuration,
with asymmetrically biased side-gates, our results are equally extendable to
other gate configurations, and possibly even to other geometries.

This paper is organized as follows: In Section~\ref{sec:system} we describe
our system and our numerical methods. Section~\ref{sec_results} is divided in
four parts. In the first part (\ref{subDispersion}) we analyze the dispersion
relations of the leads, and in the second (\ref{subModes}) we show how guided
states look in real space. A scheme for electron guiding is presented in the
third part (\ref{subGuiding}), and we use this scheme to analytically
calculate the bend and the Hall resistances in the last subsection
(\ref{subLB}). Our conclusions are given in Sec.~\ref{sec_conclusion}.

\section{System and Methods\label{sec:system}}

The studied system is shown in Fig.~\ref{fig_system}, it is a graphene cross
with four side gates ($G_1$, $G_2$, $G_3$, $G_4$) placed between four
orthogonal leads. When biased, the gates create a local potential at the
system edges, which decreases towards the interior of the system. We model the
potential of a single side gate by a Gaussian function
\begin{equation}
\label{eq_pot}
  U_g(r_n)=U_0 \exp(-r_n^2/2\sigma^2),
\end{equation}
where $r_n$ is the minimal distance from the present point to the system edge
(see Fig.~\ref{fig_system}). The width of the potential $\sigma$ is set to 10
nm, so that potentials of neighbouring gates do not overlap. We use $G_1$ as
a reference gate, and potentials of all other gates ($G_2$, $G_3$, and $G_4$)
are set opposite to that of the $G_1$, as shown in Fig.~\ref{fig_system}.


For our numerical calculations we use KWANT, a software package for quantum
transport simulations based on the tight-binding model.\cite{kwant} Graphene
is modeled in KWANT using the tight-binding Hamiltonian
\begin{equation}
\mathbf{H} = \sum_{\langle i, j \rangle} (\tilde{t}_{ij}
                   \hat{c}_{i}^\dagger \hat{c}_{j} + H.c.) +
\sum_{i}U_i \hat{c}_{i}^\dagger \hat{c}_{j},
\end{equation}
where $\hat{c}^\dagger_i$ and $\hat{c}_i$ are the electron creation and
annihilation operators, and $U_i$ is the value of the total gate potential
($U_i = \sum_{g=1}^{4} U_g(x_i, y_i)$) on the $i$-th carbon atom. The hopping
term $\tilde{t}_{ij} = t e^{i\varphi_{ij}}$ is defined using the electron
hopping energy \mbox{$t = -2.7$ eV}, and the Peierls phase factor
\begin{equation}
    \varphi_{ij} = \frac{e}{\hbar}
    \int_{\vec{r}_j}^{\vec{r}_i} \vec{A}(\vec{r}) d\vec{r},
\end{equation}
where $\vec{A}(\vec{r})$ is the vector potential. The vector potential in the
horizontal leads is set using the Landau gauge
\mbox{$\vec{A}_H=-By\vec{e_x}$}, and that in the vertical leads is
\mbox{$\vec{A}_V=Bx\vec{e_y}$}. These two potentials are smoothly connected in
the main scattering region using the procedure described in
Refs.~\onlinecite{baranger_bf, shevtsov}.

\section{Results\label{sec_results}}
\subsection{Dispersion relations}\label{subDispersion}
\begin{figure*}[t]
\includegraphics{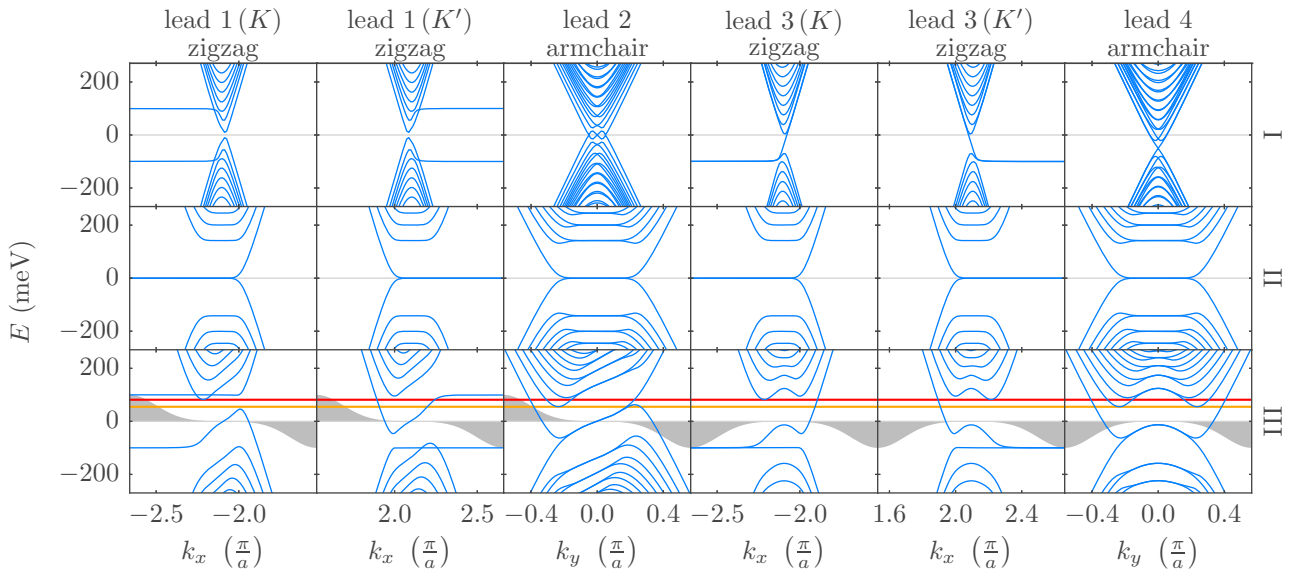}
\caption{\label{fig_dispersions}
   (Color-online) Lead dispersion relations (columns) for different
   combinations of magnetic field and side-gate potential (rows):
   \mbox{$B = 0$\,T}, \mbox{$U_0 = 100$\,meV} (case I, first row),
   \mbox{$B = 20$\,T}, \mbox{$U_0 = 0$\,meV} (case II, second row), and
   \mbox{$B = 20$\,T}, \mbox{$U_0 = 100$\,meV} (case III, third row).
   Gray horizontal lines mark the zero energy, while gray areas in the
   third row show the lead potential profile along the positive
   $x$ and $y$ directions. Since KWANT produces dispersions along the
   lead direction, we inverted the dispersions of the first and the
   fourth lead, because they point in negative $x$ and $y$ directions. The
   red line in the third row marks the minimal energy of the first band
   in the zigzag leads --- compare it with the minimal energy of the first 
   band in the armchair leads (orange line).}
\end{figure*}

First we study the dispersion relations (presented in
Fig.~\ref{fig_dispersions}) of the side-gated graphene leads. We compare three
cases with different combinations of the side-gate potential and magnetic
field. Dispersion relations without magnetic field or external potential were
extensively studied in Refs.~\onlinecite{nakada_gnr, brey_gnr}, and therefore
we do not present them here. Dispersions for a nonzero potential \mbox{($U_0 =
100\,\textrm{meV}$)} and without magnetic field are shown in the first row
in Fig.~\ref{fig_dispersions}. In our previous work~\cite{petrovic_gnr} on
quantum point contacts we investigated the dispersion relations of
symmetrically gated leads, where the same Gaussian potential as given by
Eq.~(\ref{eq_pot}) was used. Here, for zigzag leads, we focus on a narrow
wave-vector range in close proximity of the two valleys ($K$ and $K'$). The
potential on the edges determines the energy of the dispersionless bands. In
case of a symmetric potential, as in the $3^{\rm rd}$ lead, dispersionless
bands shift in energy to a value of $-U_0$. On the other hand, in
asymmetrically gated $1^{\rm st}$ lead, side-gates open a small energy gap
between the two flat bands. The gap energy is determined by the lead width.
In armchair leads, the asymmetric potential in the $2^{\rm nd}$ lead preserves
the electron-hole symmetry, while the symmetric potential in the $4^{\rm th}$
lead moves the Dirac point towards negative energies.


Results for a nonzero magnetic field and no gate potential (second row in
Fig.~\ref{fig_dispersions}) are explained in Ref.~\onlinecite{brey_qhe}. In
this case, both armchair and zigzag leads show dispersionless surface states,
appearing exactly at the energy of the Landau levels (LLs). In this regime,
graphene exhibits specific quantization of the Hall resistance, as it was
measured in Refs.~\onlinecite{graphene_qhe_zhang, novoselov_qhe}.


The most relevant case for us is when both magnetic field and side-gate
potential are present in the system (third row in
Fig.~\ref{fig_dispersions}). First noticeable difference introduced by the
side-gates is a twisting of the otherwise flat bands of the surface states
(compare the second and the third row in Fig.~\ref{fig_dispersions}). As a
consequence of this twisting, surface states become dispersed, and new states
appear in the bulk. As we show below these states appear only in specific
areas of the sample. In general the symmetry of the lead potential is
reflected in the lead dispersion. We plot the potential profiles of each lead
in the third row of Fig.~\ref{fig_dispersions} (gray areas) to show this
connection. Asymmetrically gated leads (the $1^{\rm st}$ and the $2^{\rm nd}$
lead) have asymmetric dispersions, while symmetrically gated leads (the
$3^{\rm rd}$ and the $4^{\rm th}$ lead) have symmetric dispersions. For
symmetrically gated leads this connection can be explained in the following
way: suppose we are interested in how the dispersion relation changes when,
instead along the lead, we look in the opposite direction (towards the system).
To do this we have to invert the potential profile relative to the middle
line of the lead. In the inverse space (the space of wave vectors $k$), this
view change is equivalent to inverting the dispersion relative to the \mbox{$k
= 0$} axis (all $k$ values go to $-k$, and the opposite). If the lead
potential is symmetric, then this change of view has no effect. We would
obtain the same potential and the same dispersion relation. In other words,
the inverted dispersion is equal to the initial one \mbox{$E_n(-k) = E_n(k)$}.
This is the case with the $3^{\rm rd}$ and the $4^{\rm th}$ lead.


For asymmetrically gated leads, this connection between the lead potential and
the dispersion is not so straightforward. If magnetic field is sufficiently
strong, the states moving along the opposite edges are completely separated.
These opposite edge states then feel different potentials. For example, in the
$2^{\rm nd}$ lead for zero potential, electron states with
positive velocity (and positive $k$) move along the left edge, while electrons
with negative velocities (and negative $k$), move along the right edge. For
holes, states with positive $k$ (and negative velocities) move along the left
edge, while states with negative $k$ (and positive velocities) move along the
right edge. From this we see that states with positive $k$ always move along
the left edge, while states with negative $k$ always move along the right
edge. Therefore, if we apply a potential on the left edge, it will only affect
the states with positive $k$, while if we only apply a potential on the right
edge, it will only influence the states with negative $k$. Assuming that an
electron state with positive $k$ is shifted in energy (due to the side gate
potential) by some value $\Delta E$, then electrons with negative $k$ are
shifted by -$\Delta E$, as well as hole states with the same negative $k$.
From here, we see that the dispersion is asymmetric $E_n(k) = -E_{-n}(-k)$.
This explanation is similar with the one given in chapter IV in
Ref.~\onlinecite{data}.


Before we proceed to the next part, we would like to stress one very important
difference between armchair and zigzag leads. Although the minimal band
energies appear to be similar for all leads, they are not precisely equal.
The minimal band energy in armchair leads is slightly smaller than in the
zigzag leads. The red and orange lines in the last row in
Fig.~\ref{fig_dispersions} show this small misalignment. This difference
introduces additional complexity in the system, since a new mode can open in
one lead, but electrons can not travel to the neighbouring lead, since there are
no open states there. Further below, we explain the importance of this
misalignment in more detail.

\subsection{Incoming and outgoing modes\label{subModes}}

\begin{figure}[t]
\includegraphics{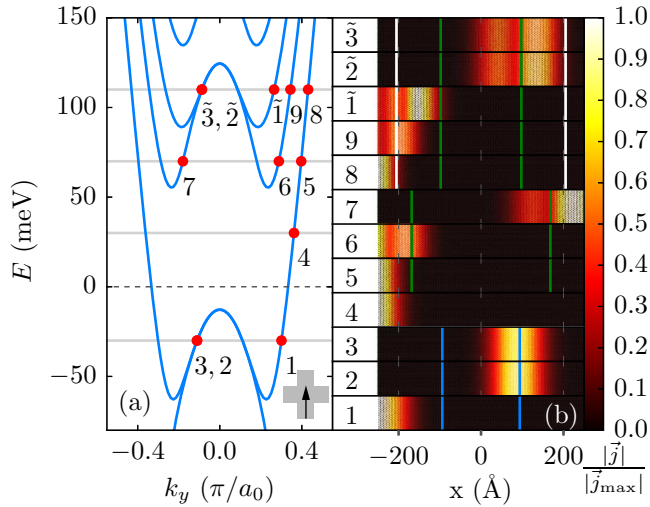}
\caption{\label{fig_guiding1}
  (Color online) (a) Dispersion relation of the symmetrically gated $4^{\rm
  th}$ lead. (b) Normalized current density across the lead for
  incoming modes marked with red circles in (a). Figure (b) is divided
  vertically (1--$\tilde 3$), with each inset corresponding to one state marked
  in (a). The considered Fermi energies are \mbox{$E_F = -30$}, $30$, $70$,
  and $110$\,meV (horizontal gray lines in (a)). The blue lines in
  (1--$\tilde{3}$) mark the equipotential lines where \mbox{$E_F - U_0 =
  E_{LL=0}$}. Similarly, green and white lines mark the positions where
  previous difference is $E_{LL=1}$, and $E_{LL=2}$, respectively. Magnetic
  field is \mbox{$B = 20$\,T} and \mbox{$U_0 = 100$\,meV}. The direction of
  the modes and the considered lead are sketched in the lower right corner in
  (a).} 
\end{figure}


To better understand the motion of charged particles in the system, 
here we analyze the incoming and outgoing modes of two representative leads:
one with symmetric (the $4^{\rm th}$ lead), and one with asymmetric (the
$2^{\rm nd}$ lead) side-gate potential. We focus on studying the evolution of
the current profiles in each lead with increaseing Fermi energy.


In Fig.~\ref{fig_guiding1}{(b)}, we present the current density profiles (insets
from 1 to $\tilde{3}$ on the right side of Fig.~\ref{fig_guiding1}) for the
incoming modes (red circles in Fig.~\ref{fig_guiding1}{(a)}). As we previously
mentioned, because of the potential symmetry, the dispersion relation is also
symmetric. Therefore the current density profiles for the outgoing modes can
be obtained by inverting the profiles shown in Fig.~\ref{fig_guiding1} along
the middle line of the lead ($x = 0$). We can differentiate two groups of
outgoing states in Figs~\ref{fig_guiding1}{(a)} and~\ref{fig_guiding1}{(b)}:
(1) the normal edge states (e.g.~states 1, 4, and 9), and (2) the guided
states (e.g.~states $\tilde{2}$, and $\tilde{3}$) which move along the
specific equipotential lines. The electron kinetic energy along these
equipotential lines match the energy of Landau levels \mbox{$E_F - U(x, y) =
E_{LL}$}. For symmetric potential, two of these equipotential lines appear on
the system edges for each new appearing LL, and with increasing Fermi energy
these lines move towards the middle of the lead \mbox{($x = 0$)}. If the
applied potential is smaller than the energy difference between two
neighbouring LLs, then a pair of these equipotential lines appear
simultaneously for each {LL}. For example, in Fig.~\ref{fig_guiding1}, the
energy difference between the $1^{\rm st}$ LL and the $2^{\rm nd}$ LL is
smaller than the applied side-gate potential. Therefore, the equipotential
lines for the $1^{\rm st}$ and the $2^{\rm nd}$ LL coexist at higher Fermi
energies (green and white lines in insets 8--$\tilde{3}$ in
Fig.~\ref{fig_guiding1}{(b)}). Because of the symmetry of the side-gate
potential, these equipotential lines always appear in pairs: the line on the
right side correspond to guided electrons coming out of the lead, while the
line on the left corresponds to guided electrons coming into the lead. Each of
these lines can accommodate two states coming from different valleys (in
armchair leads this is not so obvious, because there are no separate valleys,
but in zigzag leads each guided state can be connected with a specific
valley). For the zeroth LL, the guided states are always centered on the
equipotential line (states going along the blue lines in insets 2, and 3 in
Fig.~\ref{fig_guiding1}{(b)}), while for higher LLs there is significant broadening
of the guided states (states going along the green lines in insets
$\tilde{2}$, and $\tilde{3}$ in Fig.~\ref{fig_guiding1}{(b)}). Similar
behaviour was reported in Ref.~\onlinecite{lagasse}.


\begin{figure}[t]
\begin{center}
\includegraphics[width=0.49\textwidth]{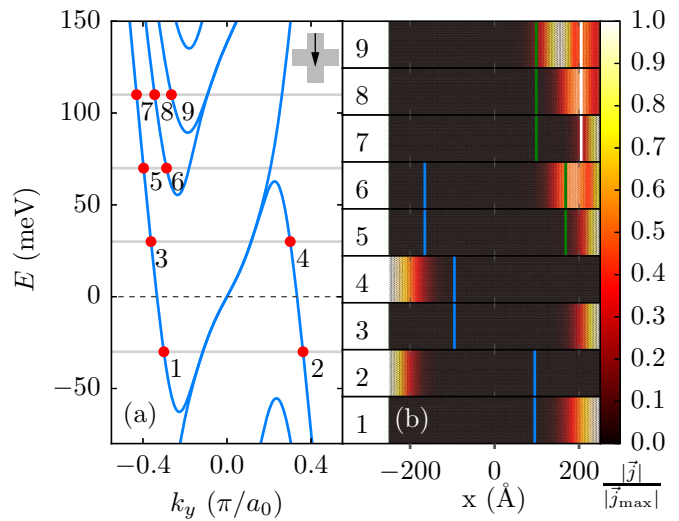}
\end{center}
\caption{\label{fig_guiding2}
  (Color online) Same as Fig.~\ref{fig_guiding1}, but now for incoming
  modes from the $2^{\rm nd}$ (asymmetrically gated) lead. Magnetic field is
  \mbox{$B = 20$\,T} and \mbox{$U_0 = 100$\,meV}.}
\end{figure}


\begin{figure}[bht]
\begin{center}
\includegraphics[width=0.49\textwidth]{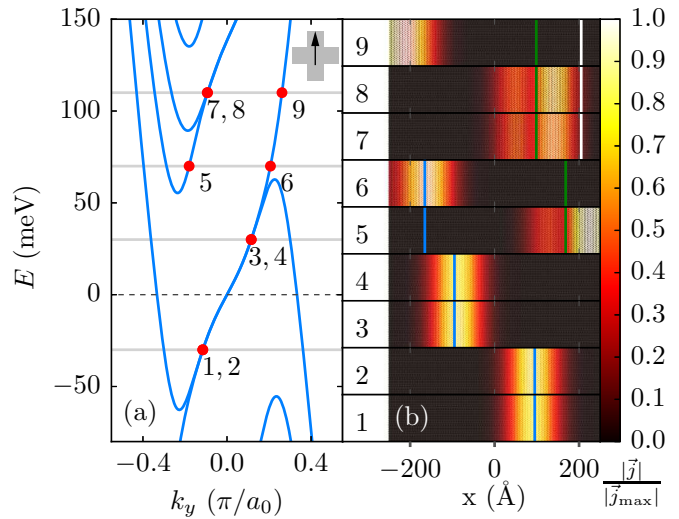}
\end{center}
\caption{\label{fig_guiding3}
  (Color online) Same as Figs.~\ref{fig_guiding1}, and~\ref{fig_guiding2},
  but now for outgoing modes in the $2^{\rm nd}$ lead. Magnetic field is \mbox{$B =
  20$\,T} and \mbox{$U_0 = 100$\,meV}.} 
\end{figure}


The case for an asymmetric potential is presented in Figs.~\ref{fig_guiding2},
and~\ref{fig_guiding3}. Here, because of the asymmetry, the outgoing modes are
not equivalent to the incoming ones, and therefore it is necessary to study
them separately. In contrast to symmetrically gated leads, here for each LL
there is only one equipotential line satisfying the condition \mbox{$E_F
- U_0 = E_{LL}$}. For each new LL this line appears first on the right edge,
and moves towards the left edge as we increase the Fermi energy (see the blue
lines in insets 1--6 in Fig.~\ref{fig_guiding2}{(b)}). For the zeroth LL,
these blue lines mark the separation point between electron and hole edge
states (a $pn$ border). For higher Fermi energies (\mbox{$E_F = 70$}, and
$110$\,meV), the hole edge state on the left side disappears, and as we see
below, it is replaced with an electron edge state moving in the opposite
direction. Although there are no guided states among the incoming modes in
Fig.~\ref{fig_guiding2}, they appear among the outgoing modes in
Fig.~\ref{fig_guiding3}. The electrons are guided along the equipotential
lines of the zeroth and the first LL, similarly as in the symmetric potential
case (for example, compare insets 1, 2, 5, 7, and 8 in
Fig.~\ref{fig_guiding3}{(b)}, with insets 2, 3, 7, $\tilde{2}$, $\tilde{3}$ in
Fig.~\ref{fig_guiding1}{(b)}).


Although we only considered current profiles of the armchair leads, the
corresponding incoming and outgoing modes in the (horizontal) zigzag leads are
very similar. The only difference is that in zigzag leads each guided state
can be connected with one of the valleys. The opening and closing of modes in
neighbouring leads do not occur at the same energy, because of a small
subband misalignment mentioned above. There are situations where only one of
the two guided states passes to the neighbouring lead, while the other guided
state is backscattered, because there is no open outgoing mode in the
neighbouring lead.

\subsection{Current guiding\label{subGuiding}}

\begin{figure}[bt]
\begin{center}
\includegraphics[width=0.49\textwidth]{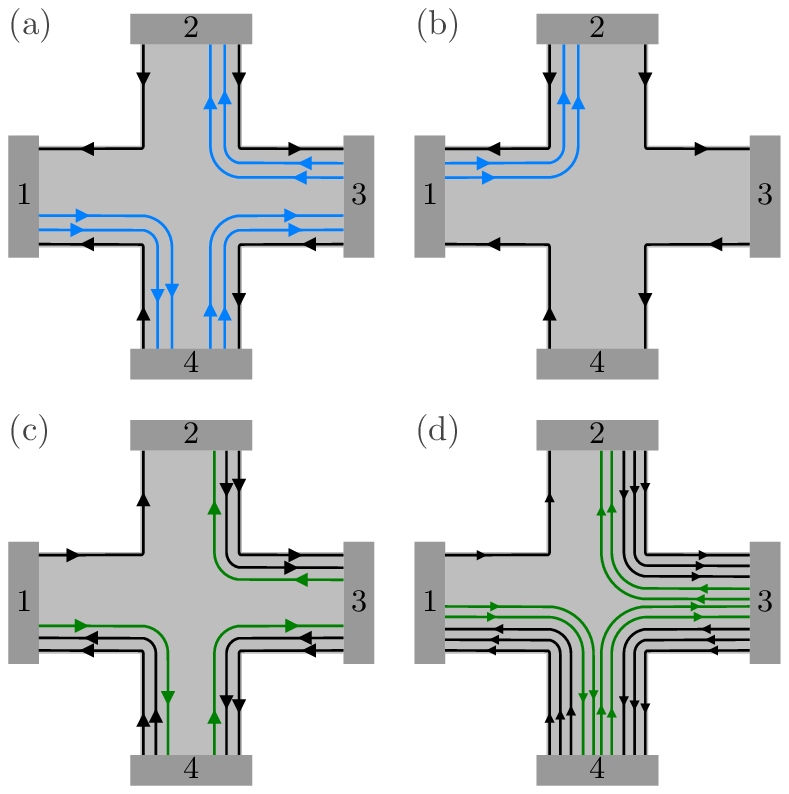}
\end{center}
\caption{\label{fig_snake}
  (Color online) Scheme of the guided states (colored lines) and the edge
  states (black lines). The corresponding energies are similar to those used
  in Figs.~\ref{fig_guiding1},~\ref{fig_guiding2}, and~\ref{fig_guiding3}: (a)
  \mbox{$E = -30$\,meV}, (b) \mbox{$E = 30$\,meV}, (c) \mbox{$E = 82$\,meV},
  and (d) \mbox{$E = 110$\,meV}. The magnetic field is \mbox{$B = 20$\,T}, and
  the side-gate potential height is \mbox{$U_0 = 100$\,meV}. Blue curves mark
  the guided states of the zeroth {LL}, while green curves mark the guided
  states of the first {LL}. The presented curves do not fully represent the
  actual current paths (they are separated from each other for better
  visualisation).}
\end{figure}


The analysis of incoming and outgoing modes allows us to construct a picture
of electron transport in this system. By extending previous results from the
vertical (armchair) leads to the horizontal (zigzag) leads, in
Fig.~\ref{fig_snake} we present a constructed scheme for electron guiding.


At the lowest energy (\mbox{$E = -30$\,meV}, Fig.~\ref{fig_snake}{(a)}), for
each edge state on the negatively biased edges, there is a pair of guided
states moving in the opposite direction. The guided states move along the $pn$
interface (the blue lines). Although there is only one $pn$ interface with
two identical guided states on it, here we show two separate blue curves in
Fig.~\ref{fig_snake}{(a)} to emphasize that there are two guided states. The
position of these lines in the scheme do not match the actual position of the
$pn$ interface. As Fermi energy increases, the $pn$ interface shifts towards
the central lines of the cross, and for positive Fermi energies, the $pn$
interface moves to the upper-left part of the system. This is what we see in
Fig.~\ref{fig_snake}{(b)}, for \mbox{$E\,=\,30$\,meV}. The two guided states
are close to the hole state on the upper-left edge. For larger Fermi energies,
the hole state on the upper-left edge disappears, and the pair of guided
states turns into a single electron edge state, moving upwards along the
upper-left edge (as in Fig~\ref{fig_snake}{(c)}).


Due to the above mentioned mismatch of the band minimal energies in
neighbouring leads, the case when the Fermi energy is \mbox{$E=70$\,meV} is
one of those situations where new modes open in the vertical (armchair) leads,
but they backscatter due to the lack of open states in the horizontal (zigzag)
leads. Therefore the scheme presented in Fig.~\ref{fig_snake}{(c)} corresponds
to larger energies (e.g~$E = 82$\,meV) when guided states open in all four
leads. This situation is very similar to that presented in
Fig.~\ref{fig_snake}{(a)}, except now there is only one guided state along the
$pn$ interface. As the Fermi energy further rises (\mbox{$E = 110$\,meV},
Fig.~\ref{fig_snake}{(d)}), a new edge state and a new guided state appear in
the system.

The scheme presented in Fig.~\ref{fig_snake} can be generalized to higher LLs,
assuming that the applied potential $U_0$ is smaller than the energy
difference between the neighbouring LLs. For $n$-th LL ($n > 0$), on the
negatively biased edges there will be $2n$ or $2n+1$ edges states, and one or
two guided states, while on the positively biased edges, there will be
\mbox{$2n - 1$} edge states. However, for every value $U_0$, no matter how
small it is, there will always be some minimal $m$ for which all higher LLs
($m' > m$) are separated by an energy smaller than the applied potential. The
present scheme is more complicated for these higher LLs, because guided states
from several LLs can coexist at the same Fermi energy. We do not consider
these cases here.

\begin{figure}[htb]
\begin{center}
\includegraphics[width=0.49\textwidth]{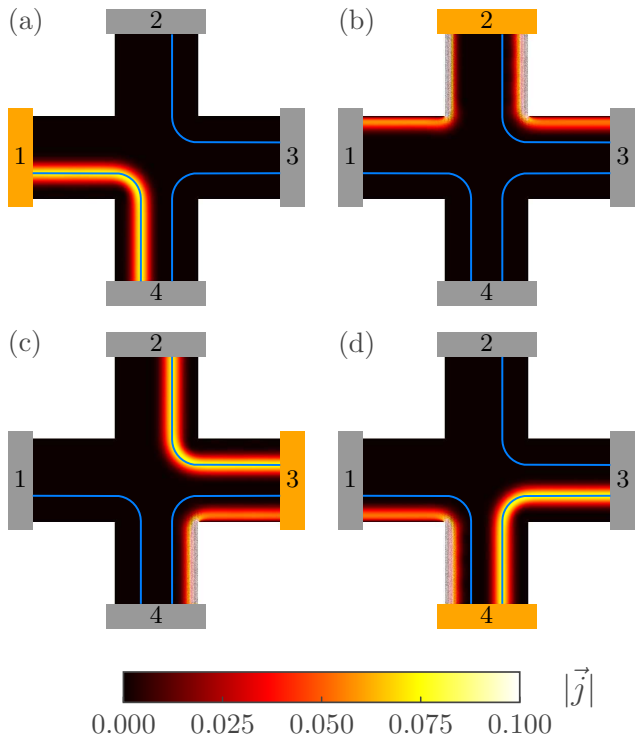}
\end{center}
\caption{\label{fig_current}
  (Color online) Current density for \mbox{$E = -30$\,meV}, \mbox{$B =
  20$\,T}, and \mbox{$U_0 = 100$\, meV}. Current is injected from the leads 
  which are colored in orange. The blue lines are equipotential lines where
  \mbox{$U(x, y)$ = $E_F$}.}
\end{figure}

To confirm the correctness of the scheme presented in Fig.~\ref{fig_snake}, we
show in Fig.~\ref{fig_current} the current density profiles for all four leads
at the Fermi energy \mbox{$E=-30$\,meV}, as obtained from our numerical
solution using {KWANT}. By combining all the currents presented in
Fig.~\ref{fig_current}, we get the same picture as that presented in
Fig.~\ref{fig_snake}{(a)}.


\subsection{Bend and Hall resistances\label{subLB}}

Based on the pictures presented in Fig.~\ref{fig_snake}, we are able
to calculate the band resistance $R_B$ by applying the Landauer-B{\"u}ttiker
formula. The most important property of the guided states
is that they fully transmit without any backscattering. In that sense, they
are equivalent to edge states. As long as there is no
backscattering, the transmission coefficients are integers and the 
transmission matrix is easy to write by hand by counting the
incoming and outgoing modes.

To calculate the resistance, we select one of the insets in
Fig.~\ref{fig_snake}, for example~\ref{fig_snake}{(d)}, and write the current
matrix
\begin{equation}
\label{eq_matrix}
\left(
  \begin{array}{c}
  I_{1} \\ I_{3} \\ I_{4}
  \end{array}
\right) = G_c
\left(\begin{array}{ccc}
  3 & 0 & -3 \\
  0 & 5 & -2 \\
  -2 & -3 & 5
  \end{array}
\right)
\left(\begin{array}{c}
  V_{1} \\ V_{3} \\ V_{4}
  \end{array}
\right).
\end{equation}
Here \mbox{$G_c = 2e^2/h$} is the conductance quantum, and we choose
\mbox{$V_2 = 0$} (for details see Chap. IV in Ref.~\onlinecite{data}).
Here we are only interested in the bend resistance
\begin{equation}
  R_{12, 34} = \frac{V_{3} - V_{4}}{I_{1}},
\end{equation}
when current is passed from the first into the second lead (the currents are
${(I_1\,0\,0)}^T$). From the second row of Eq.~(\ref{eq_matrix}) we obtain
\mbox{$V_3 = (2/5)V_4$}, and from the third row we obtain
\begin{equation}
  2V_1 = -3V_3 + 5V_4 = \frac{19}{5}V_4,
\end{equation}
and therefore \mbox{$V_1 = (19/10)V_4$}. Substituting this back in the first
row in Eq.~(\ref{eq_matrix}), we obtain
\begin{equation}
  \frac{1}{3}\frac{I_1}{G_c} = V_1 - V_4 = \frac{9}{10}V_4,
\end{equation}
and therefore \mbox{$V_4 = (10/27)I_1/G_c$}, and
\mbox{$V_3 = (4/27)I_1/G_c$}. Finally, we can calculate the
resistance as
\begin{equation}
    \frac{V_4 - V_3}{I_1} = \frac{2}{9}\frac{1}{G_c}.
\end{equation}

In a similar manner, we can calculate the quantized resistances for the other
situation depicted in Fig.~\ref{fig_snake}. For Fig.~\ref{fig_snake}{(a)} we
obtain \mbox{$R_B = (1/4)1/G_c$}, and for Fig.~\ref{fig_snake}{(c)} we also
obtain \mbox{$R_B = (1/4)1/G_c$}. For Fig.~\ref{fig_snake}{(b)} there is only
one edge state connecting the $3^{\rm rd}$ and the $4^{\rm th}$ lead,
therefore the potential on these two leads is equal (\mbox{$R_B = 0$}).

Previous calculations can be generalized for higher LLs. Assuming that applied
potential $U_0$ is smaller than the separation between the neighbouring LLs,
we can differentiate two cases. In the first case, there is only one guided
state open in each negatively biased lead (equivalent to
Fig.~\ref{fig_snake}{(c)}), while in the second case there are two such guided
states (equivalent to Fig.~\ref{fig_snake}{(d)}). In the first case, the
general current-voltage matrix relation
\begin{equation}
\label{eq_matrix_general1}
\left(
  \begin{array}{c}
  I_{1} \\ I_{3} \\ I_{4}
  \end{array}
\right) = G_c
\left(\begin{array}{ccc}
  2n & 0 & -2n \\
  0 & 2n+1 & -1 \\
  -1 & -2n & 2n+1
  \end{array}
\right)
\left(\begin{array}{c}
  V_{1} \\ V_{3} \\ V_{4}
  \end{array}
\right),
\end{equation}
can be expressed in terms of LL index $n$. When solved, this gives the
quantized resistances \begin{equation}
R_B = \frac{1}{4n^2}\frac{1}{G_c}.
\end{equation}
For the second case, when both guided states are present in the system, the
general Landauer-B{\"u}ttiker matrix is
\begin{equation}
\label{eq_matrix_general2}
\begin{pmatrix}
  I_{1} \\ I_{3} \\ I_{4}
  \end{pmatrix}
  = G_c
\begin{pmatrix}
  2n+1 & 0 & -2n-1 \\
  0 & 2n+3 & -2 \\
  -2 & -2n-1 & 2n+3
\end{pmatrix}
\begin{pmatrix}
  V_{1} \\ V_{3} \\ V_{4}
\end{pmatrix},
\end{equation}
which gives
\begin{equation}
  R_B = \frac{2}{{(2n + 1)}^2}\frac{1}{G_c}.
\end{equation}

\begin{figure}[t]
\begin{center}
\includegraphics[width=0.49\textwidth]{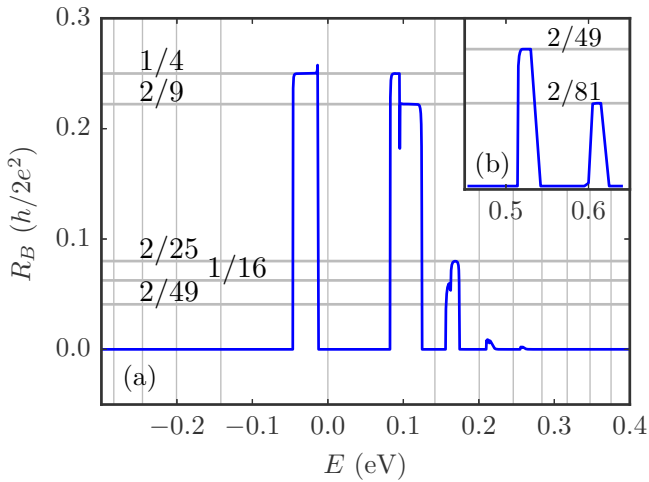}
\end{center}
\caption{\label{fig:rb}
  (Color online) (a) Bend resistance ($R_{B}$, blue curve) obtained using
  KWANT, and the analytical solutions obtained from the current matrix (gray
  horizontal lines). Gray vertical lines mark the positions LLs for \mbox{$B =
  20$\,T}. The side-gate potential is set to $U_0 = 100$\,meV. (b) Bend
  resistance for $n = 3$, and $n = 4$ Landau levels at higher fields $B =
  100$\,T, $U_0 = 50$\,meV.}
\end{figure}

Comparison between numerical and analytical results is presented in
Fig.~\ref{fig:rb}. Quantized resistances obtained analytically agree well with
the ones obtained numerically, at least for the first three Landau levels
in Fig.~\ref{fig:rb}{(a)}. For higher Landau levels, the match is not exact
(see for example the line for $R_B = (2/49)1/G_c$ in Fig.~\ref{fig:rb}{(a)}).
We suspect that the reason for this mismatch is a spatial widening of the guided
states for higher LLs, which might lead to some backscattering. A comparison
with the band resistance obtained for higher field and weaker gate potential
in Fig.~\ref{fig:rb}{(b)} reveals that the calculated resistance still matches
the analytically obtained fractional values. Also the stronger field appears
to better align the minimal band energies, since we do not observe the
generalized quantized values $R_B = (1/4n^2) 1/G_c$, where only one guided
state is present in the system. Another characteristic of $R_B$ is that is 
not symmetric for electrons and holes. Plateaus appear only for zero and 
positive LLs. A narrow positive peak near the right corner of the first 
plateau, and a negative peak between the 1/4 and 2/9 plateaus originate from
a small misalignment of subbands in the horizontal leads. Although gates $G_2$
and $G_3$ induce equal potential on the lower edge in the first and the 
third lead (see Fig.~\ref{fig_system}), this potential is slightly modified 
by gates on the upper edges (gates $G_1$ and $G_2$). Subbands are misaligned
because of this small potential difference on the lower edge. 

\begin{figure}[bth]
\begin{center}
\includegraphics[width=0.49\textwidth]{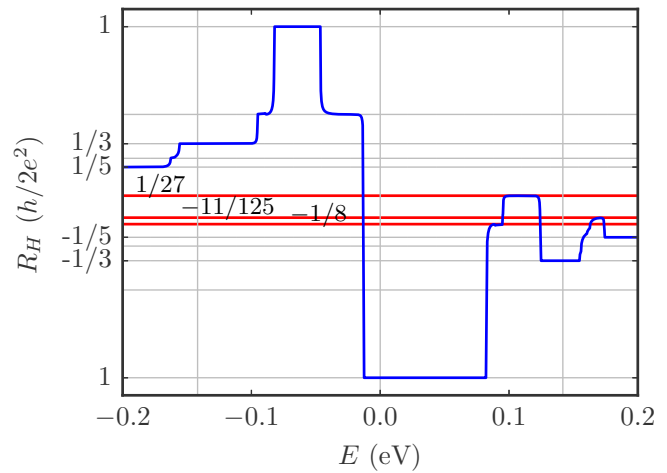}
\end{center}
\caption{\label{figRH} (Color online) 
  Hall resistance ($R_H$, blue curves) obtained using {KWANT}, and the
  analytical solutions obtained from the current matrix (red horizontal lines).
  Magnetic field and side potential are the same as in
  Fig.~\ref{fig:rb}{(a)}.} 
\end{figure}

Results for the Hall resistance are presented in Fig.~\ref{figRH}, for the
same magnetic field and side-gate potential as in Fig.~\ref{fig:rb}{(a)}.
Under the same conditions as in the case of the bend resistance, we can
calculate the quantization values for the Hall resistance. The conductance
matrix is the same as given by Eqs.~(\ref{eq_matrix_general1}),
and~(\ref{eq_matrix_general2}). But now the current is injected in the first
lead and collected in the third lead (the current column is $\begin{pmatrix}
I_1&-I_1&0\end{pmatrix}^T$). The Hall resistance is calculated analytically
as $R_{13,42} = R_H = V_4 / I_1$, because $V_2 = 0$. For the two cases, we 
obtain 
\begin{equation}
  R_H = \frac{-4n^2 + 2n + 1}{{(2n)}^3}\frac{1}{G_c},
\end{equation}
and
\begin{equation}
  R_H = \frac{-4n^2 + 5}{{(2n + 1)}^3}\frac{1}{G_c}.
\end{equation}
In Fig.~\ref{figRH}, we compare the first three analytic results 
with the numeric ones. The main feature of the
Hall resistance is that side gate potential separate the two valleys.
Instead in steps of $h/4e^2$, the plateaus are separated by $h/2e^2$ (see
horizontal grey lines). 

\section{Conclusions}\label{sec_conclusion}

In conclusion, we investigated the quantum electron transport in side-gated
Hall bars in high magnetic fields. Starting from the lead dispersion relations
which reveal new states appearing in the Landau spectrum, we proceeded to
study the current density profiles of these new states in two
representative leads. Spatially, the new states are guided along 
equipotential lines where the electron kinetic energy matches the energy of a
{LL}. Due to a full transmission of these states, the transmission matrix
contains only integers and can be solved analytically. We calculated the
quantized resistances for this asymmetric gate configuration and obtained
\begin{equation}
   R_B = \frac{1}{4n^2}\frac{1}{G_c}, \nonumber{}
\end{equation}
and
\begin{equation}
   R_B = \frac{2}{{(2n+1)}^2}\frac{1}{G_c}, \nonumber{}
\end{equation}
for the bend resistance in two cases, when there is only one and when there are
two guided states. For the Hall resistance we obtain 
\begin{equation}
  R_H = \frac{-4n^2 + 2n + 1}{{(2n)}^3}\frac{1}{G_c}, \nonumber{}
\end{equation}
and
\begin{equation}
  R_H = \frac{-4n^2 + 5}{{(2n + 1)}^3}\frac{1}{G_c}. \nonumber{} 
\end{equation}
The calculated quantized resistances match the quantized resistances obtained
with the tight-binding method. Note that these results can be easily extended
to symmetrically gated Hall bars, where potential is the same on all four
gates. Also the derived pictures of electron guiding can be applied to other
geometries with a side potential, since in general for every pair of edge
states entering the system there will be a pair of guided states moving in
the opposite direction.

\section{Acknowledgements}
This work was supported by the Methusalem programme of the Flemish government.
One of us (F. M. Peeters) acknowledges correspondence with K.~Novoselov.

%

\end{document}